\documentclass[final,5p,times,twocolumn,authoryear]{elsarticle}

\usepackage{amssymb}
\usepackage{lipsum}
\usepackage{lineno}
 \usepackage{amsmath}
 \usepackage{multirow}
\usepackage{graphicx}
\usepackage{txfonts}
\usepackage{natbib}
\usepackage{txfonts}
 \usepackage{rotating}
\usepackage{graphicx}
\usepackage{txfonts}
\usepackage{natbib}
\usepackage{txfonts}
\usepackage[bookmarks=false]{hyperref}

\usepackage{changes}

\usepackage{multirow} 
\usepackage{soulutf8}

\journal{High Energy Astrophysics}

\begin{document}

\begin{frontmatter}



\title{Evidence for 3XMM J185246.6+003317 as a massive magnetar with a low magnetic field}


\author[a]{Rafael C. R. de Lima,}
\author[b,c]{Jonas P. Pereira,}
\author[c,d]{Jaziel G. Coelho,}
\author[d,e]{Rafael C. Nunes,}
\author[d,f]{Paulo E. Stecchini,}
\author[g]{Manuel Castro,}
\author[a]{Pierre Gomes,}
\author[d]{Rodrigo R. da Silva,}
\author[d]{Claudia V. Rodrigues,}
\author[d]{José C. N. de Araujo,}
\author[b,h]{Micha{\l} Bejger,}
\author[b]{Pawe{\l} Haensel,}
\author[b]{J. Leszek Zdunik}

\affiliation[a]{organization={Universidade do Estado de Santa Catarina}, city={Joinville}, postcode={89219-710},  state={SC},country={Brazil}}
\affiliation[b]{organization={Nicolaus Copernicus Astronomical Center, Polish Academy of Sciences},  city={Warsaw},  postcode={00-716}, country={Poland}}
\affiliation[c]{organization={Núcleo de Astrofísica e Cosmologia (Cosmo-Ufes) \& Departamento de Física, Universidade Federal do Espírito Santo}, city={Vitória}, postcode={29075-910}, state={ES}, country={Brazil}}
\affiliation[d]{organization={Divisão de Astrofísica, Instituto Nacional de Pesquisas Espaciais}, city={São José dos Campos}, postcode={12227-010}, state={SP}, country={Brazil}}
\affiliation[e]{organization={Instituto de Física, Universidade Federal do Rio Grande do Sul}, city={Porto Alegre}, postcode={91501-970}, state={RS}, country={Brazil}}
\affiliation[f]{organization={Instituto de Astronomia, Geofísica e Ciências Atmosféricas, Universidade de São Paulo}, ciy={São Paulo}, postcode={05508–900}, state={SP}, country={Brazil}}
\affiliation[g]{organization={RECOD.ai, Institute of Computing, University of Campinas}, city={Campinas}, postcode={13083-852}, state={SP}, country={Brazil}}
\affiliation[h]{organization={INFN Sezione di Ferrara}, city={Ferrara}, postcode={44122}, country={Italy}}

\begin{abstract}
3XMM J185246.6+003317 is a transient magnetar located in the vicinity of the supernova remnant Kes\,79. So far, observations have only set upper limits to its surface magnetic field and spindown, and there is no estimate for its mass and radius. Using ray-tracing modelling and Bayesian inference for the analysis of several light curves spanning a period of around three weeks, we have found that it may be one of the most massive neutron stars to date. In addition, our analysis suggests a multipolar magnetic field structure with a subcritical field strength and a carbon atmosphere composition. Due to the time-resolution limitation of the available light curves, we estimate the surface magnetic field and the mass to be $\log_{10} (B/{\rm G}) = 11.89^{+0.19}_{-0.93}$ and $M=2.09^{+0.16}_{-0.09}$~$M_{\odot}$ at $1\sigma$ confidence level, while the radius is estimated to be $R=12.02^{+1.44}_{-1.42}$ km at $2\sigma$ confidence level. They were verified by simulations, i.e., data injections with known model parameters, and their subsequent recovery. The best-fitting model has three small hot spots, two of them in the southern hemisphere. These are, however, just first estimates and conclusions, based on a simple ray-tracing model with anisotropic emission; we also estimate the impact of modelling on the parameter uncertainties and the relevant phenomena on which to focus in more precise analyses. We interpret the above best-fitting results as due to accretion of supernova layers/interstellar medium onto 3XMM J185246.6+003317 leading to burying and a subsequent re-emergence of the magnetic field, and a carbon atmosphere being formed possibly due to hydrogen/helium diffusive nuclear burning. Finally, we briefly discuss some consequences of our findings for superdense matter constraints.
\end{abstract}







\end{frontmatter}




\section{Introduction} \label{sec:intro}

The X-ray pulsar 3XMM~J185246.6+003317 (hereafter 3XMM~J1852+0033) was discovered in the field-of-view of an {\it XMM-Newton} observation of the supernova remnant Kes\,79, which hosts a central compact object (CCO)~\citep{2003ApJ...584..414S}. 
These observations were independently analysed by \cite{2014ApJ...781L..16Z} and \cite{2014ApJ...781L..17R}, who reported a bright point-like source, located 7.4$'$ away from the CCO just outside the southern boundary of Kes\,79, having a very prominent periodic modulation of $P{\approx}11.6$~s. 
The source is increasing its period at a rate $\dot{P}<1.4\times 10^{-13}$~s/s, and its
\mbox{X-ray} luminosity is higher than its spin-down luminosity, ruling out a rotation-powered nature~\citep[see, e.g.,][]{2017A&A...599A..87C}. This implies a surface dipolar magnetic field strength $B<4.1\times10^{13}$~G, and a characteristic age $\tau_{\mathrm{age}}> 1.3$\,Myr. The foreground absorption ($N_H$) toward 3XMM~J1852+0033 is similar to that
of Kes\,79, suggesting a similar distance of ${\sim}7.1$ kpc~\citep[see][for details]{2004ApJ...605..742S,2014ApJ...781L..16Z,2016ApJ...831..192Z}. 3XMM~J1852+0033 has been
classified within the Soft Gamma Repeaters
(SGRs) and the Anomalous X-ray Pulsars (AXPs) class, usually called magnetars, which are neutron stars (NSs) characterized by a quiescent soft X-ray ($2-10$ keV) luminosity of the order of $10^{30}-10^{35}$ erg/s, spin period in the range $2-12$ s, and a spindown rate from $10^{-15}$ to $10^{-10}$~s/s \citep[see, e.g.,][]{2014ApJS..212....6O,2015RPPh...78k6901T,2017ARA&A..55..261K}. In particular, the low dipolar magnetic field inferred from the spindown suggests that this source is a transient magnetar with low-B \citep{2014ApJ...781L..17R}. 

Since magnetars are usually isolated NSs, inferring their macroscopic properties is a complicated task; it is usually assumed that they have a canonical mass of $1.4\,M_{\odot}$. However, the existence of high-mass NSs is also well-known (in binaries): \mbox{PSR J1614--2230} has a mass $M=1.97\pm0.04$~$M_\odot$~\citep{2010Natur.467.1081D}; the mass of 
PSR J0348+0432 is $2.01\pm0.04$~$M_\odot$~\citep{2013Sci...340..448A}, and PSR J0740+6620 has an estimated mass of $2.08\pm0.07$~$M_\odot$
~\citep{2021ApJ...915L..12F}. Observationally, the probability distribution for known NS masses presents a two component Gaussian mixture model, with mean values around $1.34\,M_{\odot}$ and $1.8\,M_{\odot}$, and standard deviations of ${\sim}0.1\,M_{\odot}$ \citep[see][for details]{2018MNRAS.478.1377A}. 

This paper reports on our analysis of the {\it XMM-Newton} X-ray data of 3XMM~J1852+0033 using ray-tracing modelling. 
We assume that its X-ray pulse profile is due to the emission of hot spots on its surface. Our main motivations are: 
(i) magnetars are usually slowly rotating NSs in which ray-tracing models are simple and relativistic phenomena are important; (ii) as far as we are aware of, there are no inferences of masses and radii of magnetars by means of ray-tracing techniques.

Ray-tracing modeling, as explored in recent studies, is pivotal for simulating X-ray light curves and understanding NS physics. This technique accounts for general relativistic effects, crucial for depicting the propagation of light near these compact objects. By incorporating detailed models of hot spots on the NS surface, ray-tracing allows for the accurate reproduction of observed pulse profiles. These simulations are instrumental in constraining NS physical parameters, such as mass and radius, by matching theoretical predictions with observational data. Studies like those of \citep{2002ApJ...566L..85B}, \citep{2013ApJ...768..147T}, and \citep{2020ApJ...889..165D}, \citep{Riley_2019}, highlight the method's utility in revealing the complex interplay between NS magnetic fields, surface temperature distributions, and geometric factors.
Therefore, ray-tracing allows us to learn more about the physics taking place around magnetars (and any other class of NSs) and independently complement what other phenomena/models already tell us about them.

In Sect.~\ref{sect:data} we describe aspects of the data selection and its reduction. Section \ref{sect:model} is devoted to the description of the model and the parameter estimation. The best-fitting results for the pulse profile of 3XMM~J1852+0033 are presented and discussed in Sect.~\ref{sect:results}. Finally, Sec. \ref{sec:summary} summarizes our main findings. \ref{ap:A} contains details about the pulse profile model. Further details about the atmosphere models for NSs are given in  \ref{ap:Atm}.

\section{Data selection, reduction, \bf{and preparation}} 
\label{sect:data} 

\subsection{\bf{The Observations}} 

The field around 3XMM~J1852+0033 was monitored by {\it XMM-Newton} on several occasions from 2004 to 2009.
The source entered a bright state at some time before 2009; it is unclear whether it went into a quiescent stage in 2009~\citep[see][and references therein]{2014ApJ...781L..16Z}.
We chose to retrieve five observations during the bright state, namely ObsIDs 0550670201 (2008 Sep. 19), 0550670301 (2008 Sep. 21), 0550670401 (2008 Sep. 23), 0550670501 (2008 Sep. 29) and 0550670601 (2008 Oct. 10), which we refer to as epochs A, B, C, D and E, respectively. The above choice is mainly due to data quality and due to reported characteristic timescales for significant hot spot motions. With relation to data quality, we have that for epochs away from the outburst the net source count rate is approximately one order of magnitude lower (than during it). Also, no glitches/anti-glitches have been reported in 3XMM~J1852+0033.
Regarding hot spot motions, we made use of knowledge stemming from observations of other magnetars. Careful monitoring of \mbox{SGR 1830-0645} by \textit{NICER}, for instance, has shown that characteristic timescales are on the order of a month \citep{2022ApJ...924L..27Y}, suggesting that fixed hot spots for smaller timescales would be a reasonable approximation. {\footnote {The analysis of \citep{2022ApJ...924L..27Y} suggests that the timescales for hot spot motion could be related to properties of the solid crust, which would be present in all NSs. Thus, analyses assuming fixed hot spots should not ignore that timescale.}} Hence, observations spanning no more than a couple of weeks may be combined, which increases the quantity of data to be fit, and improves the quality of the statistics.

\subsection{\bf{Science products extraction}} 

During the observations, the instrument EPIC-pn \citep{2001A&A...365L..18S} was operating in small window mode and did not have 3XMM~J1852+0033 within its field-of-view. EPIC-MOS \citep{2001A&A...365L..27T} cameras were operating in full window mode and did contain 3XMM~J1852+0033; however, for MOS\,1, the source fell into a CCD that was switched off in two of the observations. Thus, we only make use of data from MOS\,2. Standard data reduction and filtering procedures were conducted with the \textit{XMM-Newton} Science Analysis System (SAS, v.19.1.0). 

\begin{figure}
\centering
\includegraphics[width=0.9\hsize,clip]{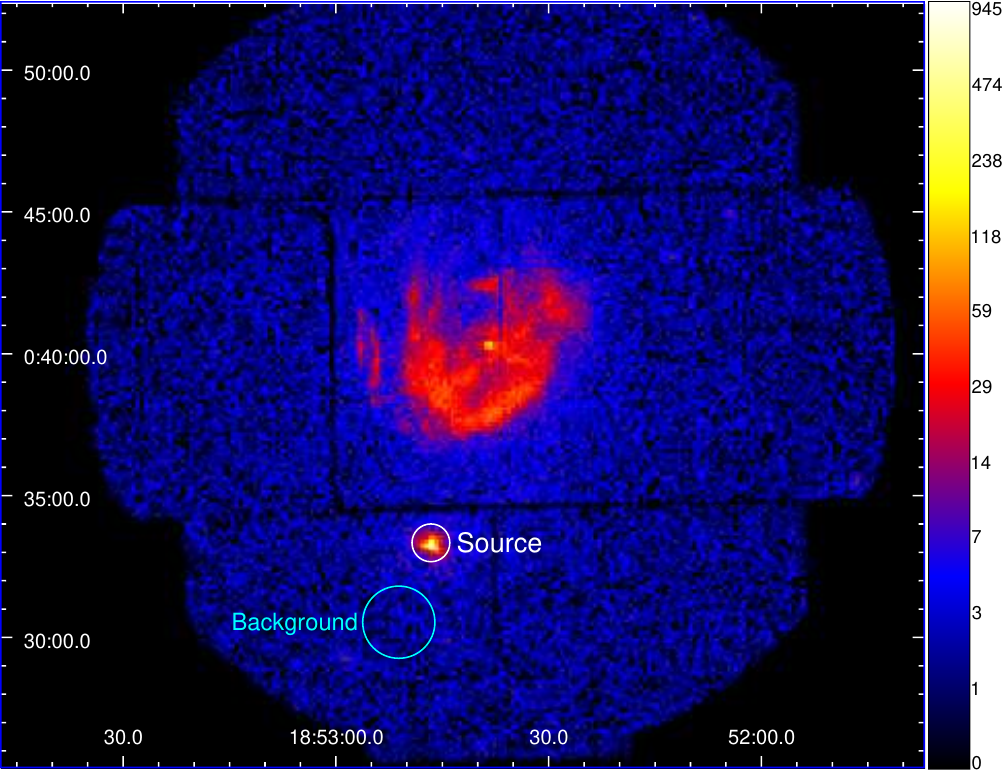}
\caption{MOS\,2 raw image for observation 0550670201. The image LUT is in log-scale; units are counts per pixel. Source and background extraction regions for 3XMM~J1852+0033 are indicated.}
\label{exposuremap}
\end{figure}

Source photons were extracted, for all observations, from circular regions of 40$''$ centred around the object's position. The background regions were chosen with the aid of the SAS task \texttt{ebkgreg}. Because this task indicates the optimal background region based solely on the detector geometry, the regions suggested were slightly shrunk to avoid the inclusion of source photons (see Figure \ref{exposuremap}).

The \texttt{barycen} and \texttt{epiclccorr} tasks were applied to extract the light curves. The former converts the photon arrival registered time into the solar system barycenter time reference and the latter performs a series of corrections to minimize effects that may impact the detection efficiency (e.g. dead time, chip gaps, point-spread-function variation\footnote{ https://heasarc.gsfc.nasa.gov/docs/xmm/sas/help/epiclccorr/}) before producing a background-subtracted light curve. The timing resolution was limited by the camera's operation mode, that is 2.6 seconds. Light curves were extracted in two energy bands: 0.3--10 and 3--8\,keV. More details are given in the next sections.

Although spectral analysis is not in the scope of this study, in order to obtain information on the source’s flux during each observation, we also extracted source and background spectra for the same aforementioned energy-band regions. Standard tasks \texttt{rmfgen} and \texttt{arfgen} were used to create the redistribution matrix file (RMF) and the ancillary response file (ARF).

\subsection{{Folded light curves production}}

We folded the light curves (pulse profiles) at the periods provided by Lomb-Scargle periodograms computed for each observations' light curves. The period values (P\,$\sim$\,11.56\,s) found for each set differ only beyond the fourth decimal place, as already shown by \cite{2014ApJ...781L..16Z} and \cite{2014ApJ...781L..17R}. They  also pointed out that there is no significant time derivative amongst observations. This is true for either energy bands (0.3--10 and 3--8\,keV). The folded light curves were binned to have 50 bins per cycle, which will be the main input to our model. For comparative testing, we also produced folded light curves with 16 bins per cycle.

To derive the flux emitted by 3XMM~J1852+0033 during each observation, we fitted the 0.3--10\,keV spectra with a simple absorbed blackbody model (\texttt{phabs*bbody}). We used the X-ray spectral fitting package \textsc{XSPEC} \citep{1996ASPC..101...17A}, version 12.11.1. The overall values of the equivalent hydrogen column ($N_{\text{H}}$\,$\sim$\,1.2--1.5\,$\times$\,10$^{22}$\,cm$^{-2}$) and blackbody temperature ($kT$\,$\sim$\,700--800\,eV) provided by the best fits are in agreement with those reported by \cite{2014ApJ...781L..16Z}. The unabsorbed fluxes (0.3--10\,keV) for each observation were computed by the \texttt{flux} command after setting the absorption component to zero. To explore how the choice of model may affect the unabsorbed flux obtained, we added a phenomenological power-law to the previous model, i.e., the combination \texttt{phabs*(bbody+powerlaw)}. The blackbody model parameters do not change, considering the confidence range. The hydrogen column tends to assume slightly higher values (from 5\% to 15\%), but it is also less constrained, agreeing within 90\% error with the blackbody model alone; the \texttt{powerlaw} index is not well constrained. These variations in the absorption component value impact the 0.3--10\,keV unabsorbed flux obtained accordingly. For further comparative testing, we have also computed the unabsorbed fluxes in the 3--8\,keV band. In this band, the role of the hydrogen column is much smaller, and the unabsorbed fluxes derived for the two models agree within 2\%.

Our modelling of the folded light curves relies on the shape of each curve, meaning that it is able to fit the curves regardless of the choice of normalization. Nevertheless, as it will be clarified in the following sections, using a sort of representative value of the source's emission, such as the unabsorbed fluxes, may help improve the quality of the results.

\section{Model description and parameter estimation} 
\label{sect:model} 

In our previous work \citep{2020ApJ...889..165D}, as well as in the original reference that introduced the method \citep{2013ApJ...768..147T}, the calculations were performed using the normalized flux, defined as $(F_{max}+F_{min})/2$. However, in the present work, we make use of a modified normalization, $(F_{max}+F_{min})/2\bar{F}$, where $\bar{F}$ denotes the mean value of the observed unabsorbed flux in the period, derived from the spectral fitting. This modification is of great significance because normalizing the theoretical pulse to the correct level improves the statistical sampling and prevents the acceptance of parameters that describe unrealistic situations, which do not correspond to the actual flux. Details are given in \ref{ap:A}.

The spacetime outside the star is well-described by the Schwarzschild metric, i.e., we neglect rotational effects, as the source is a slow rotator. We also make use of the spectral tables for models of magnetic NS atmospheres calculated by \cite{HO20071432,Ho_2008} and \cite{10.1111/j.1365-2966.2007.11663.x}, which have been made public and implemented in the \textsc{XSPEC}
package \citep[see][]{1996ASPC..101...17A,2009Natur.462...71H,ho_2013}. These neutron star atmosphere models were obtained using the most up-to-date equation of state (EOS) and opacity results for a partially ionized, strongly magnetized hydrogen or mid-Z element plasma (e.g., carbon, oxygen, and neon). The associated spectra come from the solution to the coupled radiative transfer equations for the two-photon polarization modes in a magnetized medium. The atmosphere is assumed to be in radiative and hydrostatic equilibrium. The atmosphere models depend on the NS surface temperature, the NS surface magnetic field strength $B$ and its inclination relative to the radial direction. Moreover, there is a dependence on the NS surface gravity $g = (1+z_g)GM/R^2$, where the gravitational redshift is $1+z_g = 1/\sqrt{1-2GM/Rc^2}$. More precisely, the tables used for the spectra of the neutron star atmospheres were calculated using discrete magnetic fields: $[10^{10},~10^{11},~10^{12}, ~3.0\times10^{13}]$~G (\texttt{nsmaxg}, for an H atmosphere), and $[10^{12}$, $10^{13}]$~G (\texttt{nsmaxg}, for a C atmosphere). For low magnetic fields ($B\leq 10^{10}$~G), one can use \texttt{nsx} for a non-magnetic atmosphere (H, He, and Ca). Even though we are working with a magnetar, analysis taking into account $P$ and ${\dot P}$ suggest that 3XMM~J1852+0033 is a low-B one \citep{2014ApJ...781L..17R}. In addition, 3XMM~J1852+0033's proximity to SNR Kes 79 could allow magnetic-field burying due to supernova debris; the age of the SNR Kes 79  [estimated between 4.4 and 6.7\,kyr \citep[see, e.g.,][]{2016ApJ...831..192Z,2022ApJ...928...89H}] and its proximity  also suggest that 3XMM~J1852+0033's surface burial is recent enough so field re-emergence is still an ongoing process. All the above means that NS atmosphere tables with subcritical $B$s are reasonable.
The tables also depend on discrete values of the surface gravity $g$. For a C atmosphere, we adopted $\log_{10}(g ~[\rm{cm/s^2}])=[13.6-15.3]$, and for a H atmosphere, $\log_{10}(g~[\rm{cm/s^2}])=[13.6-15]$. To be able to consider continuous values of the model parameters in the Bayesian analysis, the model emission has been interpolated using the tabulated emission values.
Further details about the atmosphere models for NSs are given in \ref{ap:Atm}.

We used the Markov Chain Monte Carlo (MCMC) method to determine the set of parameters $\bar{\theta}_i$ that best describes the 3XMM~J1852+0033 light curves. For the case of three spots, which is the maximum number of spots we considered in the modelling, they are

\begin{equation}
\begin{aligned}
 \bar{\theta}_i= \Big\{ & M,\,R,\,\log(B/G),\,i,\,\theta_{1},\,\theta_{2},\,\phi_{2},\,\theta_{3},\,\phi_{3}, \\
 & \theta_{c1},\,T_1,\,\theta_{c2},\,T_2,\,\theta_{c3},\,T_3 \Big\}, 
\end{aligned}
\label{eq_parameters}
\end{equation} 
with the parameters separated into two categories: the {\it fixed} and the {\it secular} parameters. The fixed parameters are presented in the first row and do not change for the epochs investigated (NS mass $M$ and radius $R$, its magnetic field $B$, position of the hot spots $\theta_i$ and $\phi_i$ $(i=1,2,3)$ and, for reference, $\phi_1\equiv 0$). The secular parameters may have different  values for each epoch (sizes and temperatures of hot spots, $\theta_{ci}$ and $T_i$). Table \ref{table_prior} summarizes their definitions and prior range distributions. The posterior probability distribution function is  

\begin{equation}
 p({\cal D}\mid\bar{\theta}) \propto \exp \Big( - \frac{1}{2} \chi^2\Big) \, ,
\end{equation}
with
\begin{equation}
\chi^2 = \sum_{i=1}^N (F^{obs} - F^{th}(\bar{\theta}_i))^{T} \Sigma^{-1}(F^{obs} - F^{th}(\bar{\theta}_i)), 
\end{equation}
where $F^{obs}$ represents the observed flux as detailed in  \ref{ap:A}, $F^{th}$ denotes the total theoretical flux generated by the hot spots, and $\Sigma^{-1}$ is defined as the inverse of the square N-by-N covariance matrix, which is associated with the variability of the observed flux. The goal of any MCMC approach is to draw $N$ samples $\bar{\theta}_i$ from the general posterior probability density
\begin{equation}
\label{psd}
p(\bar{\theta}_i, \alpha\mid{\cal D}) = \frac{1}{Z} p(\bar{\theta},\alpha) p({\cal D}\mid\bar{\theta},\alpha)  \, ,
\end{equation}
where $p(\bar{\theta},\alpha)$ and $p({\cal D}\mid\bar{\theta},\alpha)$ 
are the prior distribution and the likelihood function, respectively. Here, ${\cal D}$ and $\alpha$ are the set of observations and possible nuisance parameters and $Z$ is the normalization factor.

We perform the statistical analysis based on the \texttt{emcee} algorithm \citep{2013PASP..125..306F}, assuming the theoretical model described in \ref{ap:A} and the uniform priors summarized in Table \ref{table_prior}. 
During our analysis, we discarded the first 20\% steps of the chain as burn-in. We follow the Gelman-Rubin convergence criterion \citep{10.1214/ss/1177011136}, checking that all parameters in our chains have ${\cal R} - 1 < 0.01$, with ${
\cal R}$ quantifying the Gelman-Rubin statistics, also known as the potential scale reduction factor. 

In order to better quantify the agreement (and/or disagreement) among models, we have adopted the Akaike information criterion (AIC) \citep{Akaike}, defined as
\begin{equation}
\text{AIC} \equiv -2 \ln  \mathcal{L}_{\rm max} + 2p,
 \end{equation}
where $ \mathcal{L}_{\rm max}$ is the maximum likelihood value and $p$ is the total number of free parameters in the model. 
The AIC is derived from an approximate minimization of the Kullback-Leibler divergence \citep{10.1214/aoms/1177729694} between the data-fit distribution and the true distribution.
For statistical comparison, the AIC difference between the model under study and the reference model is calculated. 
It has been argued in \cite{Tan_2011} that a model is preferred over another one if their AIC difference is larger than a threshold value $\Delta_{\rm thr}$. As a rule of thumb, $\Delta_{\rm thr} = 5$ can be considered the minimum value to assert a strong support in favour of the model with a smaller AIC value, regardless of the properties of the models under comparison \citep{Liddle_2007}.

\begin{table*}
\caption{Ranges of the model parameters. The table shows the parameters used in the model, which are divided into fixed parameters, whose values remain constant between epochs, and secular parameters, which can evolve over time. The magnetic field range depends on the chosen atmosphere, such as H ($10^{10}-10^{13}$ G), O ($10^{12}-10^{13}$ G), and C ($10^{12}-10^{13}$ G). Although extrapolations beyond this range may be allowed for numerical purposes, any walker chains outside the acceptable interval are discarded to ensure full coverage.}
\centering 
\begin{tabular}{c c c c c c c c c c c c} 
\hline
	  \multicolumn{3}{c}{{\bf{Fixed parameters}}}\\
		 \hline
	 $M(\text{M}_{\odot}$) & stellar mass & $0.1-2.3$ \\
		 $R$(km) & stellar radius & $7.5-20.0$  \\
		 $T$~(keV) &  stellar surface temperature & $0.0-0.2$ \\
		 $\log_{10}(B/G)$   & magnetic field strength & $10.0-13.0$  \\
		 $i$ & angle between the LOS and the rotation axis  & $0-90^\circ$\\
		 $\theta_1$ & spot one's colatitude & $0-90^\circ$\\
		 $\theta_2$ & spot two's colatitude  & $0-90^\circ$\\
        $\phi_2$ & spot two's longitude & $0-360^\circ$ \\
          $\theta_{3}$ &  spot three's colatitude & $0-180^\circ$   \\
		  $\phi_3$ & spot three's longitude & $0-360^\circ$ \\
		 \hline
		  \multicolumn{3}{c}{{\bf{Secular parameters}}}\\
		 \hline
		  $\theta_{c1}$ & spot one's  semi-aperture  & $0-90^\circ$\\
		 $T_{1}$~(keV) &  spot one's temperature & $0.0-1.0$ \\
		 $\theta_{c2}$ &  spot two's semi-aperture & $0-90^\circ$   \\
		 $T_{2}$~(keV) &  spot two's temperature & $0.0-1.0$ \\
		 $\theta_{c3}$ & spot three's  semi-aperture  & $0-90^\circ$\\
		  $T_{3}$~(keV) &  spot three's temperature & $0.0-1.0$ \\
		  \hline
\end{tabular}
\label{table_prior}
\end{table*} 

\begin{figure}
\centering
\includegraphics[width=1.0\hsize,clip]{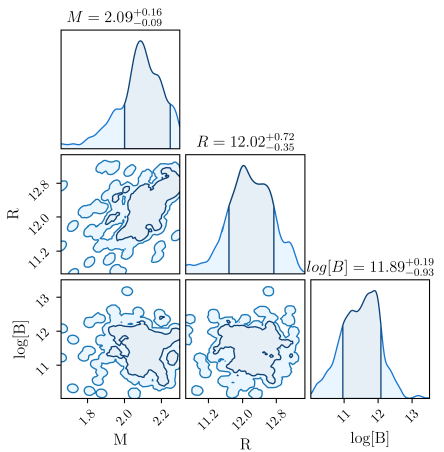}
\caption{Parametric space at 1$\sigma$ (dark blue) and $2\sigma$  (light blue) confidence levels (CLs) and one-dimensional marginalized distribution of $M$, $R$ and $\log(B)$ for three hot spots on the surface of 3XMM~J1852+0033.  For completeness, $M$, $R$ and $B$ are measured in units of $M_\odot$, km, and $G$, respectively. The quoted values of the mass, radius and surface magnetic field are at 1$\sigma$ CL. 
Our control simulations indicate that the radius of 3XMM~J1852+0033 is reliable at $2\sigma$ CL $\left(R=12.02^{+1.44}_{-1.42} \rm{km} \right)$, while its mass and surface magnetic field are reliable at $1\sigma$ CL.
}
\label{PS_plot}
\end{figure}

\begin{figure}
\centering
\includegraphics[width=1.0\hsize,clip]{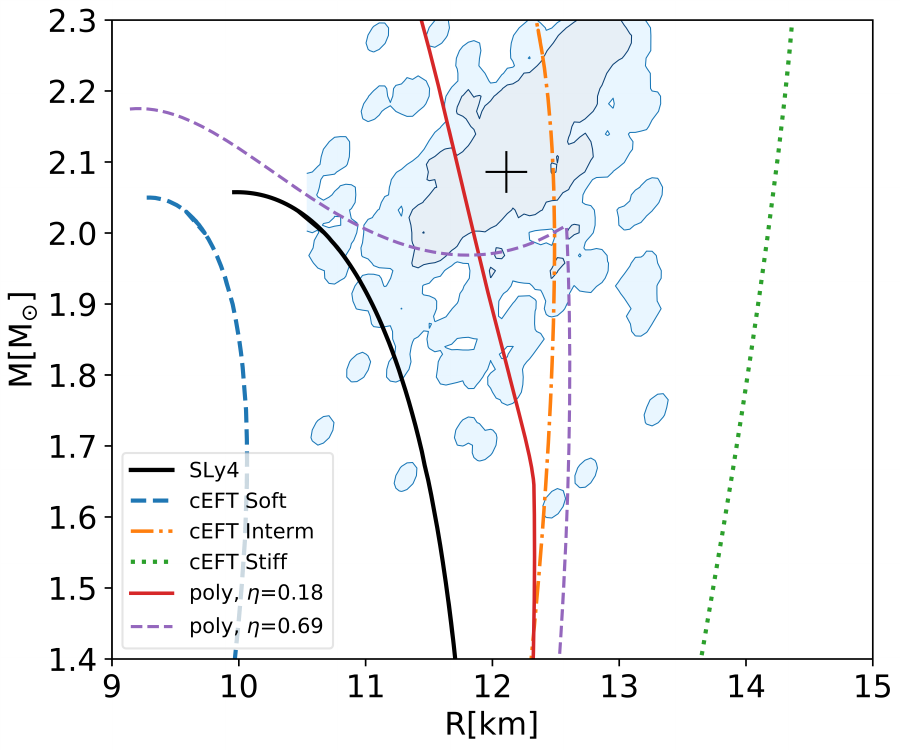}
\caption{$M-R$ relations for (purely hadronic) soft, intermediate and stiff cEFT EOS models and also for the SLy4 EOS. For completeness, selected polytropic EOS models for hybrid stars with different energy density jumps ($\eta$ values) are presented (for details on them, see the main text).
The cusps in the associated curves mark the phase transition masses beyond which an exotic (e.g., quark) phase appears. The dark blue and the light blue areas are the posterior constraints for 1$\sigma$ (68$\%$) and 2$\sigma$ ($95\%$) CL, respectively. The black cross marks the median of our best-fitting stellar parameters.}
\label{EOS_canditates}
\end{figure}

In order to minimise the effects from the light curve's time resolution, as well as from the binning choice when building the folded light curves (pulse profiles), two precautions were taken prior to the fitting. First, our pulse profile model was calculated taking into account the data time resolution, i.e., the model was convoluted with a top-hat function with 2.6\,s width.  Secondly, synthetic pulse profiles with different binning were produced to test 
if after convergence -- according to our Bayesian analysis -- the injection data could be recovered.
It turns out that the synthetic pulse profiles produced with the same time resolution of our folded light curve data allow recovery of the injected mass and surface magnetic field of the star within $1\sigma$ ($68\%$) CL. The injected radius of the star, though, could only be retrieved at $2\sigma$ ($95\%$) CL. The number of hot spots has been correctly recovered in all cases, while the positions and temperatures of the spots only within $95\%$ CL. Thus, to be conservative given our data quality,
we only interpret radius related observables, geometrical properties and temperatures of hot spots at $2\sigma $ CL.

\section{Results and Discussion} 
\label{sect:results} 

\begin{figure*}
\centering
\includegraphics[width=0.45\hsize,clip]{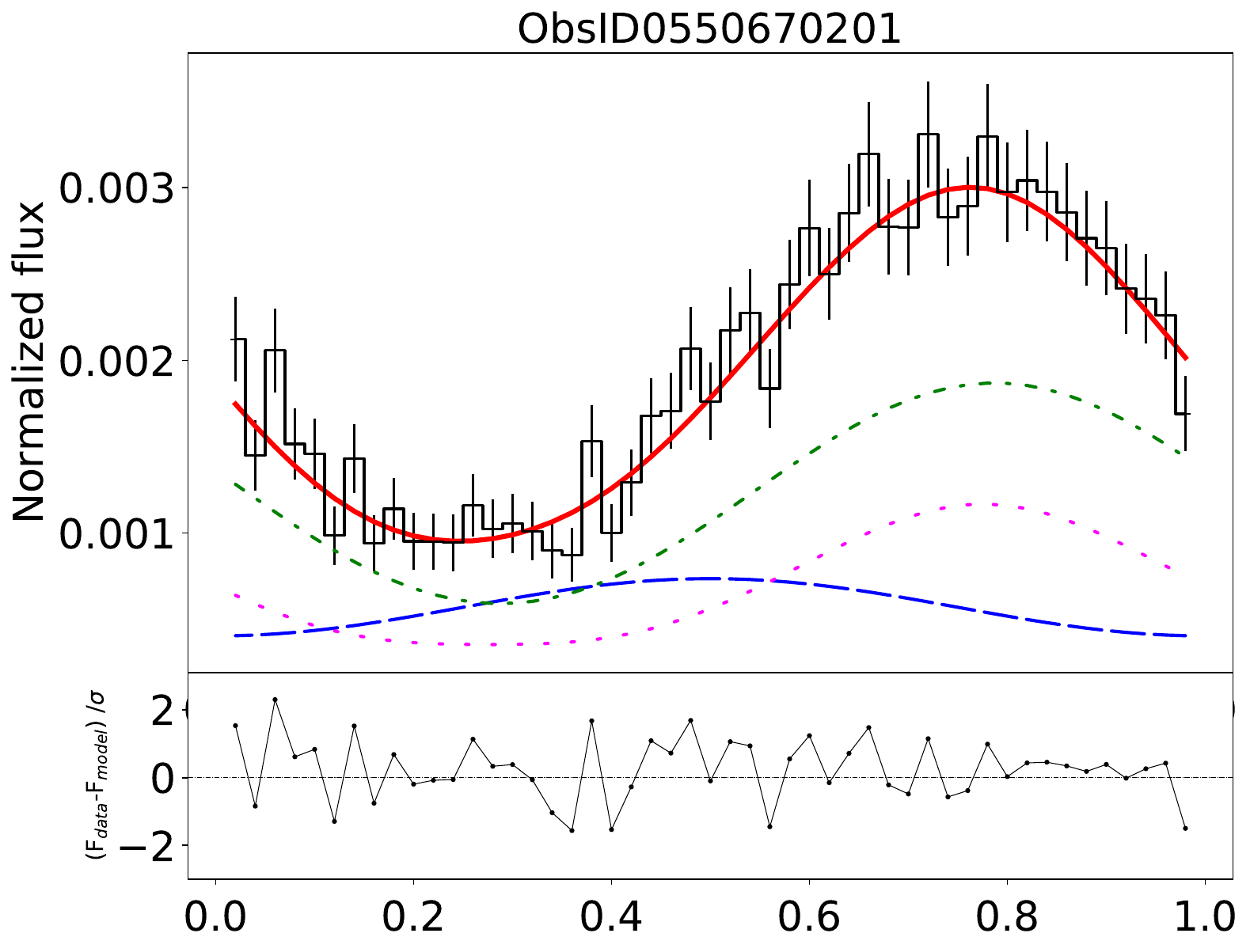} 
\includegraphics[width=0.45\hsize,clip]{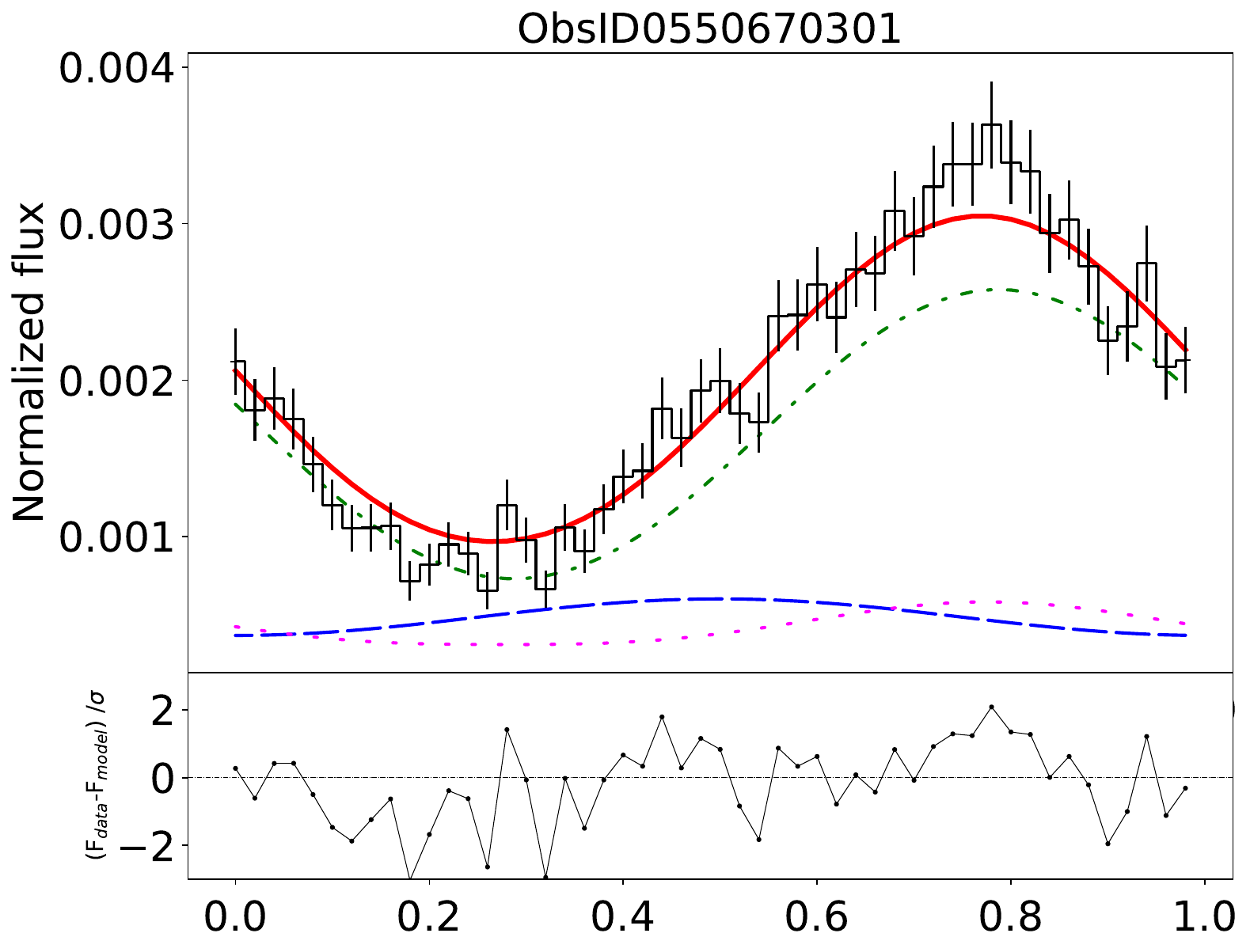} 
\includegraphics[width=0.45\hsize,clip]{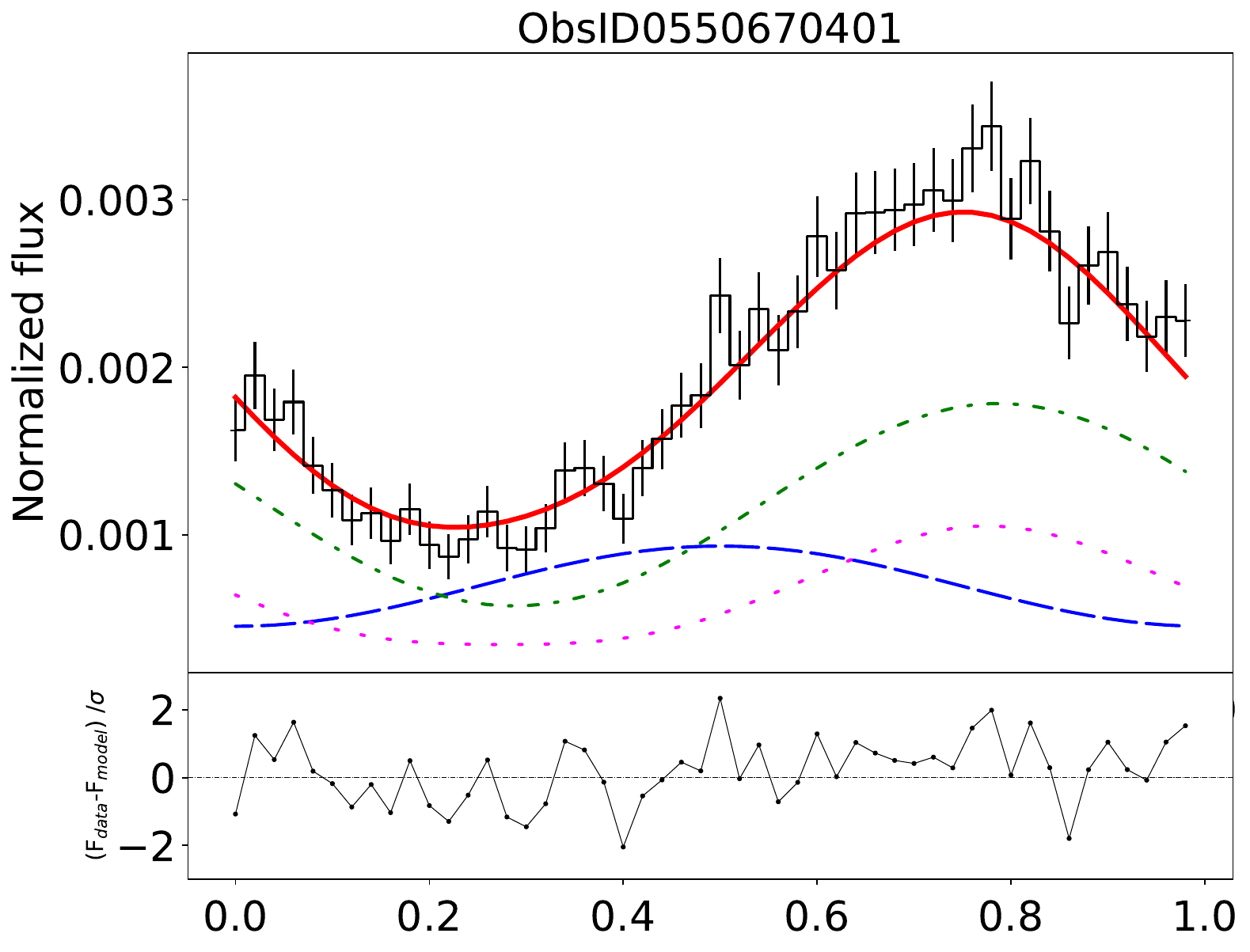} 
\includegraphics[width=0.45\hsize,clip]{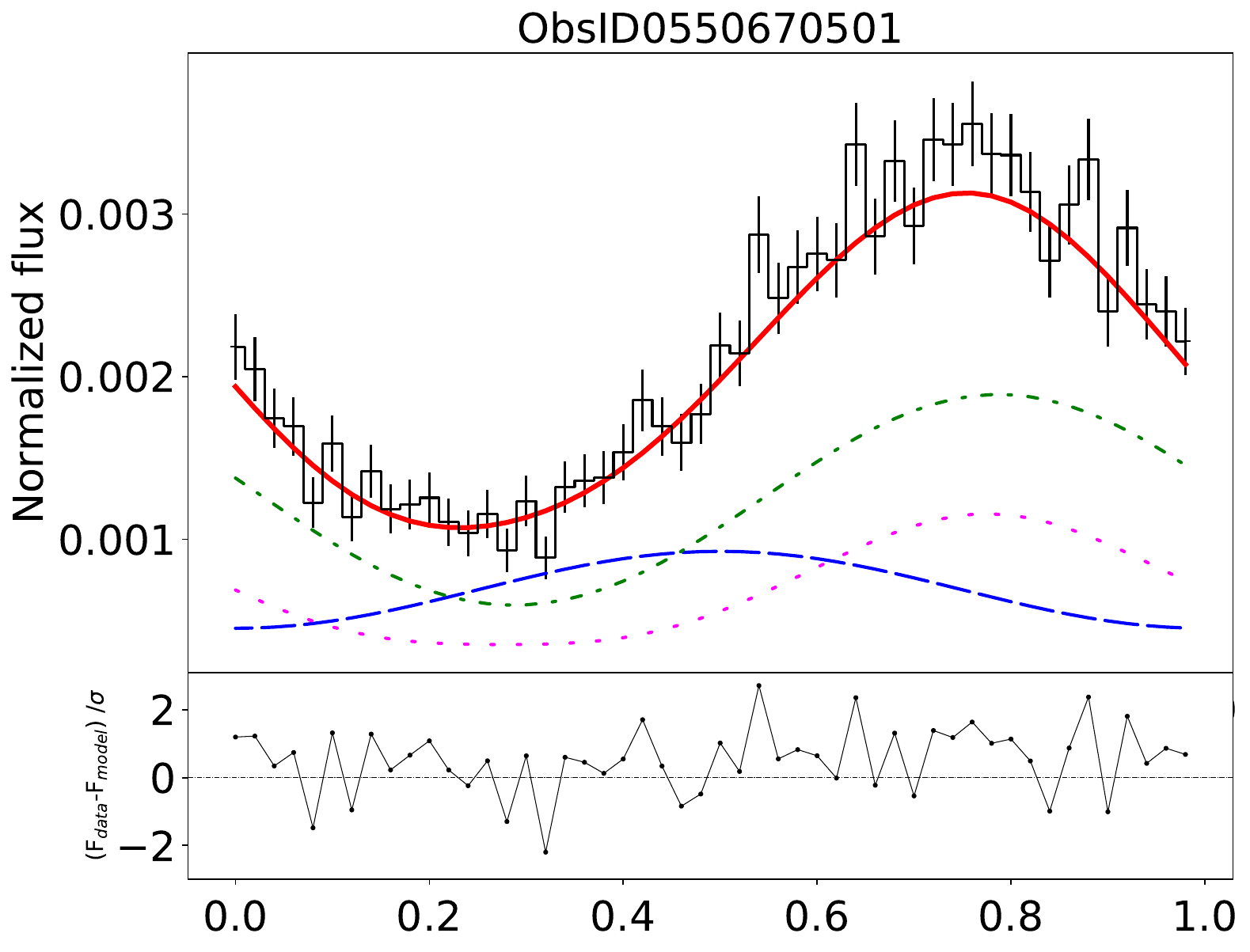}
\includegraphics[width=0.45\hsize,clip]{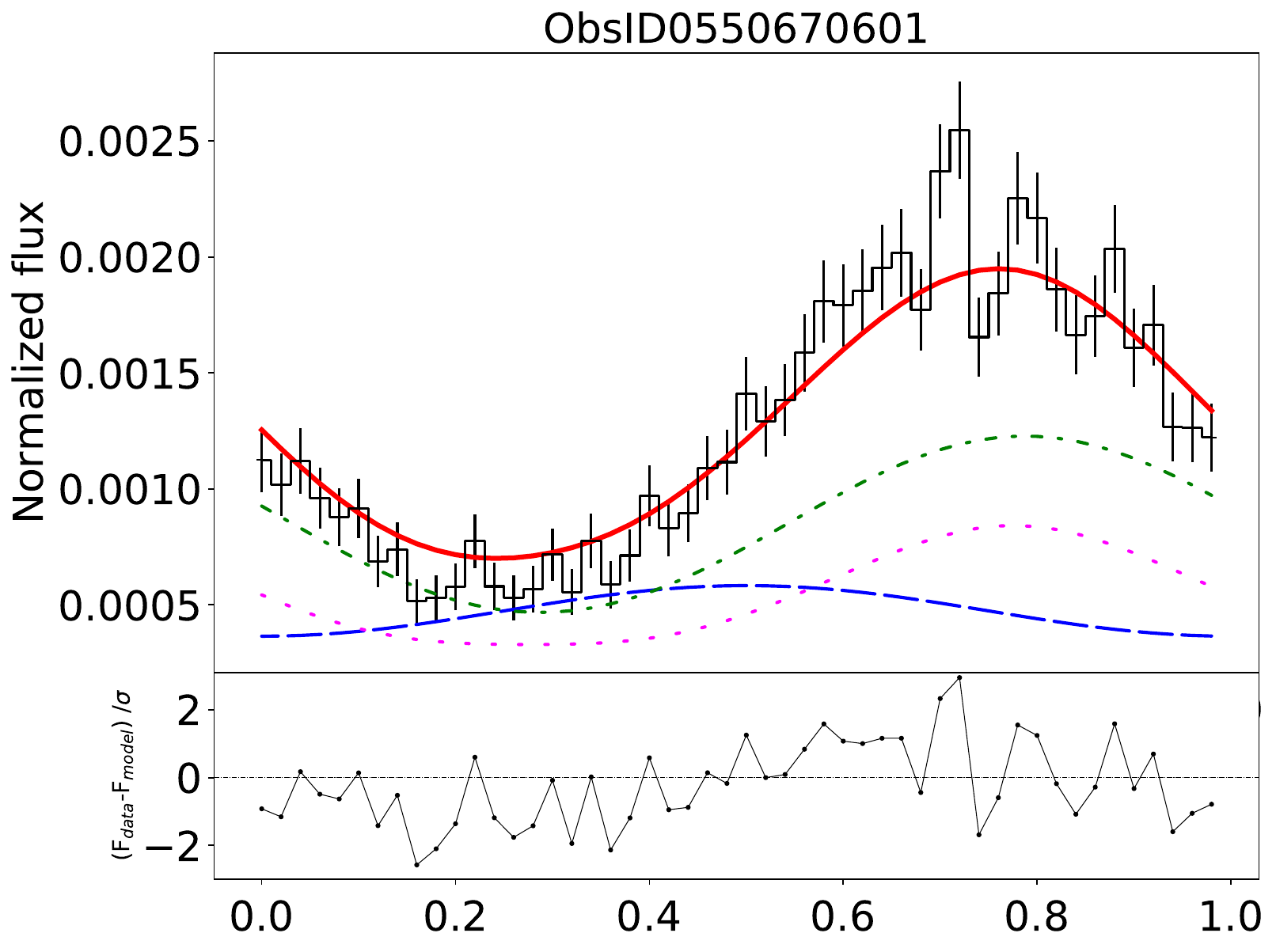} 
\caption{Phase-folded pulse profiles in counts per second for each epoch we have investigated. Observational datapoints (black histogram) and the best fits (red solid curves) are plotted. The energy band is $0.3-10$~keV. The dashed blue curves correspond to the contributions to the phase-folded pulse profiles due to spot 1 (identified by $\theta_1$), the dot-dashed green curves due to spot 2 (identified by $\theta_2$), and the dotted magenta curves due to spot 3 (identified by $\theta_3$).}
\label{fig:pulse_profile}
\end{figure*}

\begin{figure*}
\centering
\includegraphics[width=0.53\hsize,clip]{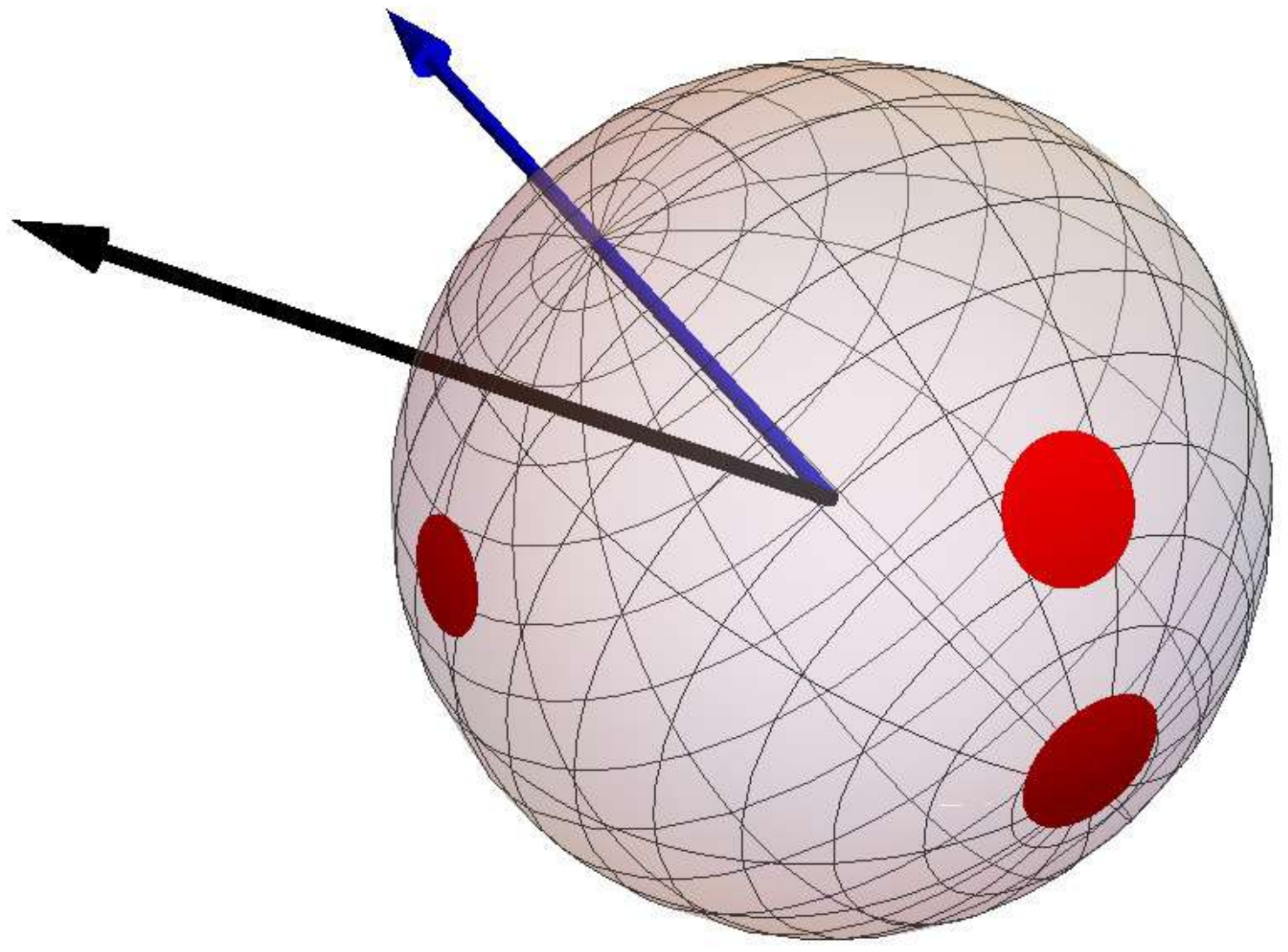}\includegraphics[width=0.34\hsize,clip]{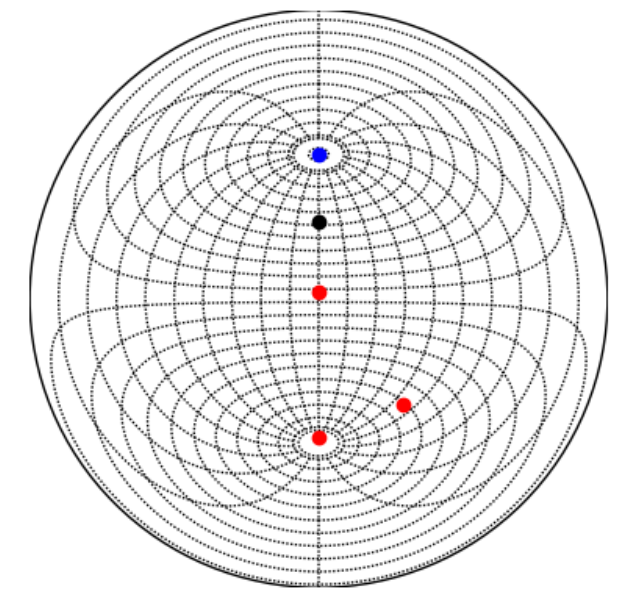}
\caption{Geometry of the best fit: 3D representation (left) and stereographic projection (right). On the left (right) are shown the spots as red circles (red dots), and the rotation axis and the line of sight with the blue arrow (blue dot) and the black arrow (black dot), respectively.
}
\label{fig:geometry}
\end{figure*}

We have investigated models with different numbers of hot spots and atmosphere compositions. Details about them can be found on Table~\ref{table:data}. The best model has been found to have three spots and a carbon atmosphere. Its AIC was 362.36 ($\chi^2/{\rm d.o.f.}= 280.36/230=1.22$), while for two hot spots, it was 647.41 ($\chi^2/{\rm d.o.f.}=581.41/242=2.40$). A model with a single spot not even presented a convergence of the chains, and it has been discarded. For a model with four hot spots, we have not obtained a decrease in $\chi^2$ larger than $1\sigma$ with respect to the three spots' case. Given this penalty, 
it is disfavoured over a model with three spots. Note that in the case of three hot spots, we are working on a 16D reference parameter space. However, the total number of free parameters depend on the number of epochs analyzed and also on the number of fixed and secular parameters. In our case of 5 epochs, 6 secular and 10 fixed parameters, one has in total $10+6\times 5=40$ free parameters. Only for reference, the total number of data-points we have for the 5 epochs is 270 (thus, ${\rm d.o.f.=230}$). The best fit to the data (done for all the epochs simultaneously) and the associated hot spot geometry on the star are shown in Figs. \ref{fig:pulse_profile} and \ref{fig:geometry}, respectively.

\begin{table*}
\centering
\caption{
{Comparison of AICs and $\chi^2$s for models with one, two, and three hot spots and different atmosphere compositions for the energy band 0.3--10keV.}}
\begin{tabular}{l c c c c c}
\hline
 Atmosphere & number of spots & AIC & $\mathbf{\chi^2}$ & d.o.f. \\
\hline
\multirow[]{3}{*}{Hydrogen} & one & 6251.00 & 7852.00 & 254 \\
& two & 1312.11 & 1902.11 & 242 \\
& three & 2735.00 & 2223.00 & 230\\
\hline
\multirow[]{3}{*}{Oxygen} & one &  3630.27  & 3540.27 & 254 \\
& two & 1630.15  & 1540.15 & 242 \\
& three & 1029.94  & 939.94 & 230\\
\hline
\multirow[]{3}{*}{Carbon} & one & 2581.00 & 2766.00 & 254 \\
& two & 647.41 & 581.41 & 242 \\
 & three & 362.36 & 280.36 & 230 \\
\hline
\end{tabular}
\label{table:data}
\end{table*}

\subsection{Best-fitting stellar parameters} 

Table \ref{tab:main_results} summarizes the main results of our statistical analysis at 1$\sigma$ (68\% CL). Throughout the text, we also give results for the 95\% CL. Due to our data quality and the results from our data injections, physical outcomes involving the radius of 3XMM~J1852+0033, are physically interpreted only at the $95\%$ CL. Figure \ref{PS_plot} summarizes the parameter space for some key NS parameters, namely $M$, $R$ and $B$. The all-epoch fit for 3XMM~J1852+0033 data suggests that $M=2.09^{+0.16}_{-0.09}$ $M_{\odot}$, $R=12.02^{+0.72}_{-0.35}$ km and $\log_{10} (B/{\rm G}) = 11.89^{+0.19}_{-0.93}$ at $1\sigma$; for $95\%$ CL, we have that $M=2.09^{+0.21}_{-0.49}$ $M_{\odot}$, $R=12.02^{+1.44}_{-1.42}$ km and $\log_{10} (B/{\rm G}) = 11.89^{+1.51}_{-1.89}$.
Our results are in full agreement with the recent NICER estimates for the equatorial radius of PSR J0740+6620 \citep[see][]{2021ApJ...918L..28M,2021ApJ...918L..27R}. In terms of compactness, ${\cal C}\equiv MG/(Rc^2)$, one has 
that $0.246\leq {\cal C}\leq 0.268$.
For the median parameters, it follows that ${\cal C}=0.256$. 

One can also assess the detectability of the NS parameters found by means of the universal relations connecting the compactness to the dimensionless tidal deformability $\Lambda$, the later obtained by the gravitational-wave (GW) detectors during the inspiral of binary NS systems \citep{2018PhRvL.121p1101A}:  
\begin{equation}
    {\cal C}=\sum_{k=0}^2a_k(\ln \Lambda)^k,
    \label{C-Lambda-relation}
\end{equation}
with $a_0=0.360$, $a_1=-0.0355$, $a_2=0.000705$ \citep{2017PhR...681....1Y}. By numerically solving Eq.~\ref{C-Lambda-relation}, one finds the associated tidal deformations for the possible combinations of masses and radii. For the median $M$ and $R$ values, it follows that $\Lambda_{{\cal C}=0.256}=23$, which is at the threshold of detectability by third generation GW detectors \citep{2022PhRvD.105h4021C}. 

We compare the results against selected purely hadronic reference equations of state: the chiral effective field theory (cEFT) models \citep[see, eg.,][]{Hebeler:2013nza,2020ApJ...901..155G}, which are in agreement with nuclear physics experiments and astrophysics, and also the benchmark nucleonic SLy4 EOS \citep{2001A&A...380..151D}. For completeness, we also use some polytropic sharp phase transition EOSs for hybrid stars as presented in \cite{2022arXiv220101217P} with different energy density jump $\eta$ values\footnote{The polytropic parameters for the EOS with $\eta=0.18$ are: $(n_0=0.475\, {\rm fm}^{-3}, n_{02}=0.413\, {\rm fm}^{-3}, \gamma_1=3.5, \gamma_2=6.0, n_{cc}=0.20\, {\rm fm}^{-3})$. The parameters for $\eta=0.69$ are: $(n_0=0.50\, {\rm fm}^{-3}, n_{02}=0.322\, {\rm fm}^{-3}, \gamma_1=3.75, \gamma_2=6.5, n_{cc}=0.20\, {\rm fm}^{-3})$. For the precise meaning of each parameter, see \cite{2022arXiv220101217P}.
}. As shown in Fig. \ref{EOS_canditates}, the intermediate stiffness chiral EOS is privileged to the detriment of (very) soft and stiff EOSs. Additionally, within $95\%$ CL, the SLy4 EOS is a viable possibility. Rotation may change the picture in general, increasing the equatorial radii and masses for any EOS \citep{2016MNRAS.459..646B}. But this is not relevant for 3XMM~J1852+0033 due to its large period (${\sim}11.6$ s); the effect of rotation on spacetime is of the order of $10^{-8}$ \citep{2007ASSL..326.....H}. More precise analysis should be performed to check for specific impact of our findings on current multi-messenger constraints for the EOS.

\begin{table*}
\begin{center}
\begin{minipage}{\textwidth}
\caption{Estimated fixed and secular parameters of 3XMM~J1852+0033 obtained from the modelling of {\it XMM-Newton} light curves.}
\label{tab:main_results}
\centering
\begin{tabular}{c c c c c c c c c c c c} 
\hline
\multicolumn{10}{c}{\textbf{Fixed parameters}} \\
 \hline
  {$M(\text{M}_{\odot}$)} & {$R(\text{km})$} & {log$(B/G)$}& {$T(\text{keV})$} & {$i$} & {$\theta_1$} & {$\theta_2$} & {$\phi_2$} & {$\theta_3$} & {$\phi_3$ }   \\ [0.5ex]
 \hline
  $2.10^{+0.16}_{-0.09}$ &
                $12.02^{+0.72}_{-0.35}$ &
                 $11.89^{+0.19}_{-0.93} $ &
                 $0.16^{+0.14}_{-0.06}$ 
                 &
                $41.77^{+0.81}_{-1.25}$ &
                 $86.10^{+1.08}_{-0.91}$ &
                $176.0^{+1.7}_{-2.3}$ &
                 $76.6^{+3.4}_{-3.0}$&
                $140.5^{+2.1}_{-2.1}$ &
                 $80.8^{+3.4}_{-4.2}$ \\ [1ex] 
 \hline
\end{tabular} 
\vspace{2ex} 
\begin{tabular}{c c c c c c c c} 
\multicolumn{8}{c}{\textbf{Secular parameters}} \\
\hline
  {Epoch} & {ObsID} & {$\theta_{c1}$} & {$T_{1}(\text{keV})$} & {$\theta_{c2}$} & {$T_{2}(\text{keV})$} & {$\theta_{c3}$} & {$T_{3}(\text{keV})$}  \\ 
 \hline
 		 A & 0550670201 & $7.75^{+0.56}_{-0.52}$ &  $0.462^{+0.024}_{-0.031}$ & $10.17^{+0.48}_{-0.41}$ & $0.801^{+0.028}_{-0.031}$ 
		  & $9.47^{+0.22}_{-0.18}$
		   & $0.779^{+0.034}_{-0.024}$\\ 
		 B & 0550670301 & $5.37^{+0.45}_{-0.49}$ &  $0.684^{+0.030}_{-0.035}$ & $12.26^{+0.46}_{-0.38}$  & $0.683^{+0.035}_{-0.011}$  
		  & $7.71^{+0.23}_{-0.19}$
		   & $0.36^{+0.018}_{-0.030}$\\ 
		 C & 0550670401 &  $7.99^{+0.45}_{-0.54}$ &  $0.550^{+0.030}_{-0.024}$ & $10.68^{+0.44}_{-0.63}$ & $0.644^{+0.024}_{-0.027}$  
		  & $9.42^{+0.18}_{-0.28}$
		   & $0.65^{+0.023}_{-0.035}$\\ 
		 D & 0550670501 &  $9.31^{+0.56}_{-0.43}$ &  $0.459^{+0.024}_{-0.028}$ & $10.24^{+0.46}_{-0.41}$ & $0.955^{+0.024}_{-0.029}$  
		  & $9.41^{+0.19}_{-0.27}$
		   & $0.958^{+0.028}_{-0.022}$\\ 
		 E & 0550670601 &  $7.78^{+0.68}_{-0.33}$ &  $0.331^{+0.029}_{-0.027}$ & $7.83^{+0.36}_{-0.45}$ & $0.905^{+0.026}_{-0.030}$ 
		  & $7.51^{+0.16}_{-0.27}$
		   & $0.897^{+0.025}_{-0.031}$\\ [1ex] 
 \hline
\end{tabular}
\footnotetext{All angles are in degrees.}
\end{minipage}
\end{center}
\end{table*}

\subsection{Magnetar with standard pulsar magnetic field} 

We have found that the best-fitting value for the magnetic field strength \mbox{($B\sim 10^{11}-10^{12}$ G)} is consistent with the magnetic field inferred from the star's spindown ($B<4.1\times 10^{13}$ G) \citep[see][]{2014ApJ...781L..17R}. However, in this latter case, only the dipolar field is taken into account. The presence of three hot spots  suggests that the magnetic field should actually be multipolar~\citep[see e.g.,][and references therein]{2020ApJ...889..165D}. The subcritical value inferred, comparable to ordinary pulsars, suggests that 3XMM~J1852+0033 has experienced burying due to accretion. Possible origins to the accreted material are supernova layers or the interstellar medium. Indeed, a small accreted mass amount (when compared to the stellar mass) of around $5\times 10^{-4}M_{\odot}$ would already be enough to bury magnetic fields of the order of $10^{14}$ G and lead to surface fields around $10^{11}-10^{12}$ G for around $10^3-10^4$ yrs \citep{2012MNRAS.425.2487V}. Considering that supernova remnant Kes\,79 can be a physically near object and that its estimated age is between 4.4 and 6.7\,kyr \citep[see, e.g.][]{2016ApJ...831..192Z,2022ApJ...928...89H}, our results suggest that we may be just seeing the re-emergence of $B$, even for multipolar field components.

\subsection{High mass and the possibility of dense-matter phase transitions} 

A relevant consequence of our best-fitting results is the possibility of a high mass for 3XMM~J1852+0033. The values obtained are in the same range of the most massive pulsars \citep{2010Natur.467.1081D,2013Sci...340..448A,2016ApJ...832..167F,2020NatAs...4...72C}, and below the estimated maximum mass (for nonrotating stars) inferred from GW and kilonova analysis (${\sim}2.16^{+0.17}_{-0.15}\,M_{\odot}$) \citep{2018ApJ...852L..25R}. Therefore, if confirmed, 3XMM~J1852+0033 would constitute one of the most massive neutron stars to date. Interestingly, the best-fitting value for $M$ using three spots is very close to the inferred mass for two hot spots (at $1\sigma$ CL, $M=2.00^{+0.19}_{-0.15}\,M_{\odot}$), meaning that a large mass persists for different models, strengthening the case of 3XMM~J1852+0033 being a massive magnetar. 

The possibility of a high mass for 3XMM~J1852+0033 also makes it natural to speculate if it may have an exotic (e.g., quark) core. Statistical studies suggest that the phase transition mass should be high, around $2M_{\odot}$ \citep{Annala:2019puf}, and this would render 3XMM~J1852+0033 another strong candidate for probing exotic cores.
For this, as a first approach, we have investigated a large sample of polytropic sharp phase transition EOSs (for further details, see \cite{2022arXiv220101217P}). 
With our posteriors, one can restrict the exotic-hadron density jumps in stars. We find, by using dozens of thousands of parametric EOSs fulfilling known astrophysical constraints that at $1\sigma$ CL, 
$\eta\equiv \Delta \rho/\rho_{\rm trans}\equiv \rho_- /\rho_+-1=0.129^{+0.211}_{-0.099}$ where $\rho_- (\rho_+)$ is the density at the top of the exotic phase (bottom of the hadronic phase). For 
$95\%$ CL, it follows that $\eta =  0.129^{+0.861}_{-0.129}$.
These results are in broad agreement with more rigorous Bayesian inferences combining multi-messenger astronomy, 
e.g., $0.40^{+0.20}_{-0.15}$ \citep{2021PhRvC.103c5802X} and 
$0.28^{+0.39}_{-0.25}$ \citep{2021ApJ...913...27L}, 
while slightly favoring weak phase transitions (small energy density jumps; see, e.g., \cite{2019A&A...622A.174S} and references therein). At $95\%$ CL, strong phase transitions are however also possible. For an example, see Fig.~\ref{EOS_canditates} ($\eta=0.69$).
Our polytropic EOSs and posteriors also allow us to estimate the phase transition mass $M_{\rm trans}$, beyond which an exotic core appears. At $1\sigma$ CL, we get 
$M_{\rm trans} = 1.39^{+0.36}_{-0.31} M_{\odot}$. Within $95\%$ CL, $M_{\rm trans} = 1.39^{+0.75}_{-0.56}M_{\odot}$. 
This means that exotic matter cores may be present in NS with masses around the Chandrasekhar mass \citep{2018MNRAS.478.1377A} all the way to massive NSs, in agreement with other statistical studies \citep{Annala:2019puf}. For comparison, we also estimate $\eta$ and $M_{\rm{trans}}$ with our set of polytropic EOSs using the inferred parameters of PSR J0030+0451 ($M=1.34^{+0.15}_{-0.14} M_{\odot}$, $R=12.71^{+1.14}_{-1.19}$ km)
\citep{2019ApJ...887L..24M,2019ApJ...887L..21R}. At $1\sigma$ CL, we have $\eta= 0.243^{+0.377}_{-0.183}$ and $M_{\rm trans} = 1.50^{+0.29}_{-0.26}M_{\odot}$. For $95\%$ CL, $\eta= 0.243^{+0.987}_{-0.240}$ and $M_{\rm trans} = 1.50^{+0.63}_{-0.48}M_{\odot}$. Clearly, 3XMM~J1852+0033 and PSR J0030+0451 (and PSR
J0740+6620) lead to consistent phase-transition parameter estimates.

\subsection{Carbon atmosphere} 

Another aspect from our best-fitting result concerns the atmosphere of 3XMM~J1852+0033. We find that the best fit points to the carbon composition, and disfavours (ionized) hydrogen or helium atmospheres. 
Blackbody models (no atmosphere and beaming) are not favored by our analysis either. While 3XMM~J1852+0033 is definitely not the CCO in the Kes\,79 supernova remnant \citep{2011MNRAS.414.2567H}, its carbon atmosphere renders it a CCO-like 
object, as it shares this feature with nearly half of the young CCOs, e.g., the youngest of them, the Cas\,A remnant \citep{2021MNRAS.506.5015H,2013A&A...556A..41K,2009Natur.462...71H}. As for CCOs, a carbon atmosphere associated with a buried magnetic field 
could be explained by the depletion of hydrogen due to its diffusion into the denser and hotter layers of a young NS, where protons are captured by heavier elements. This is the diffusive nuclear burning of hydrogen acting in young NSs with surface temperatures $T_{\rm eff}>1.5$ MK \citep[see e.g.][]{2003ApJ...585..464C,2004ApJ...605..830C}. This process generates a hydrogen current flowing downwards into the hydrogen-burning layer, and it leads eventually to the emergence of the carbon atmosphere. A similar mechanism can work for helium atmospheres, but for somewhat higher temperatures ($T_{\rm eff}>2$~MK, see \cite{2010ApJ...723..719C}). The effective  temperatures of the hot spots of 3XMM~J1852+0033, shown in Tab.~\ref{tab:main_results}, are higher than 5~MK ($0.43 {\rm keV}$), and therefore satisfy the condition for an efficient diffusive nuclear burning. Specifically, at such $T_{\rm eff}$, a hydrogen/helium atmosphere is completely burned into carbon in less than a few days \citep{2004ApJ...605..830C,2010ApJ...723..719C}. Any hydrogen/helium atmosphere after an outburst of duration of at least 200 days (the estimated span of the bright phase of 3XMM~J1852+0033) could only result from the accretion of these elements from its environment.  For completeness, carbon debris from the supernova remnant Kes 79 falling onto 3XMM~J1852+0033 might be another possibility for its carbon atmosphere. However, in this case, carbon should reemerge after accretion of hydrogen and helium from the interstellar medium, which could possibly happen via the DNB of the freshly accreted plasma.

\subsection{Impact of the modelling on the parameter uncertainties}

It may seem surprising that the uncertainties in mass and radius are only around $10\%-15\%$, given that the XMM-Newton MOS2 data has a time resolution that is only about 1/4 of the rotational period of the magnetar. However, we have taken this into account in our analysis by convolving the flux integration with a step function that has a temporal width equivalent to the satellite resolution. Importantly, our precise results are due to the use of data from five epochs and their simultaneous fit. Additionally, a considerable number of our parameters are fixed, which significantly increases the precision of our results. Finally, although the number of bins per phase is relatively low (50), we obtained precise results for the temperature and hot spot maps on the star because the pulse profiles have an approximate sinusoidal reconstruction (meaning that the precise number of points is not crucial) and because we are fitting the parameters of a specific ray-tracing model. The price to pay for this pulse-profile reconstruction is the difficulty of finding fine-structure details on the star, which we have indeed ignored. Although what we obtain are basically ``smoothed out'' results, they still tell us interesting aspects about the star. For example, models with one or two hot spots are not as good as three ones, and this can already give us interesting insights about the properties of the magnetic fields in magnetars. In addition, three hot spots are needed due to the slight deviation of the phase-folded pulse profiles from sinusoids. It is known that smaller uncertainties for the stellar parameters arise in the case of sinusoidal waveforms when hot spots are around the stellar equator \citep{2013ApJ...776...19L}, which is the case for two hot spots in our best-fitting results. This could also help explain the relatively small mass and radius uncertainties we found.

We also stress that the choice of model to compute the unabsorbed fluxes for the normalization leads to systematic uncertainties in our parameter estimations. As mentioned, these values generally agree within $10\%-15\%$ for different models in the whole band (0.3--10\,keV). As we discussed in our previous work \citep{2020ApJ...889..165D}, these systematics roughly translate into uncertainties of up to $5\%$ for the NS mass and $7\%$ for the NS radius. However, these uncertainties are still smaller than the reliable ones we estimated for the mass and radius of 3XMM J1852+0033.

In highly magnetized neutron star atmospheres, such as those of magnetars, the orientation of the magnetic field lines with respect to the star surface's normal can affect the observed flux and spectra \citep{1994A&A...289..837P,2001MNRAS.327.1081H}. In order to assess the relevance of that for small hot spots, one should look at the local fluxes \citep{1994A&A...289..837P}. Figures 2 and 3 of \citep{1994A&A...289..837P} help us estimate the expected relative changes for different angles $\vec{B}$ makes with the normal to the surface ($0$ degrees and approximately $40$ degrees) when taking the fixed value $B=4.7\times 10^{12}$ G. One roughly has that relative differences for the spectral intensity are up to $40\%-50\%$ for the peak transparency angles and magnetic field poles, and up to around $20\%-30\%$ elsewhere. (The numbers change slightly depending on the photon energies.) One would expect smaller differences for smaller magnetic field strengths, such as the ones coming from our best-fitting results. For larger hot spots, such as ours, the resultant effect on the flux of the orientation of the magnetic field with respect to the normal to the NS surface may also decrease because it should be integrated over the whole spot, which smooths out the effect \citep{2001MNRAS.327.1081H}. However, the above flux differences are not negligible in general and would also impact the mass and radius inferences by around $5\%-10\%$. Although smaller than our estimated errors for the macroscopic parameters of 3XMM J1852+0033, the general conclusion here is that more precise analysis should not ignore the angle the magnetic field makes with the NS's surface normal.

For the modelling of magnetars' hot spot emissions, beaming should not be overlooked either \citep{2019MNRAS.485.4274H}. While it should arise self-consistently from a given microscopic atmospheric model, its description is not simple in general. A way to overcome that when one wants to test several atmosphere models is to assume an empirical beaming function whose free parameters are inferred simultaneously with other stellar parameters by Bayesian analyses for each atmosphere \citep{2023arXiv230809319S}. Here, since we were interested in first estimates, we have considered the $I\propto \cos^n\alpha$ phenomenological beaming model \citep{2001ApJ...559..346D} for all possible atmospheric compositions, where $\alpha$ is the angle the ray makes with the normal to the surface (for further details, see  \ref{ap:A}). Due to the absence of accretion columns, which favor $n=2-3$ choices \citep{2001ApJ...559..346D}, we took $n=1$. Since this effective beaming model does not take into account specific atmosphere models, the carbon atmosphere that we have obtained as the best-fitting result should be seen as indicative. However, fits of free beaming parameters of NICER's PSR J0030+0451 show significant overlaps for different atmosphere compositions and little impact on the originally inferred PSR J0030+0451 parameters, based on a fully ionized hydrogen atmosphere model \citep{2023arXiv230809319S}. This suggests that in some cases the atmospheric composition does not largely affect the NS parameters and the beaming function. Indeed, we found that different choices of the beaming function, such as the Hopf one \citep{2001ApJ...559..346D} and the one associated with $n=3$, did not result in acceptable fits for the folded pulse profiles in any atmosphere case. Thus, based on all the above, uncertainties for the mass and radius of 3XMM J1852+0033 might be small when associated with variations of the (correct) beaming functions for different atmospheric compositions. We leave further investigation on this important point for future work.

\subsection{Additional tests} 
\label{sect:tests} 

In order to increase confidence in our main results, we conducted a few tests using subsets of our data. The main goal is to assess the impact of data treatment choices and model-dependent factors on the analysis. Specifically, how the binning of the pulse profile and/or the 
energy band used to extract the light curve affect the results. First, we re-performed the modelling for the same bins/phase as before (50), but for data within a more restricted energy band (from 3 to 8 keV). The lower limit chosen is so that the computation of unabsorbed flux is less sensitive between models (as mentioned, in this band the fluxes agree within $2\%$), and the upper limit is just to avoid a high background portion of the data. We briefly talk about the case with 16 bins/phase at the end of the section.

We proceeded with the Bayesian analysis exactly as in the case of the 0.3--10 keV energy band and obtained consistent results. Indeed, the best-fitting model in the 3--8 keV energy band was also found to be the one with three hot spots and a carbon atmosphere. Moreover, the inferred values for the mass, radius, and magnetic field were all in agreement (within the uncertainties) with those obtained from the broader energy band. Specifically, we obtained $2.20^{+0.10}_{-0.05}M_{\odot}$, $11.80^{+0.56}_{-0.52}$ km, and ${\rm log_{10}}(B/G)=12.05^{+0.27}_{-0.03}$ for the mass, radius, and surface magnetic field, respectively (within 1$\sigma$ CL). The difference between the results from the more restricted energy band and the broader one can be entirely explained in terms of flux uncertainties due to the modelling of unabsorbed flux (which contributes to approximately $5\%-10\%$ uncertainty in the mass and radius of 3XMM J1852+0033). Within 1$\sigma$ CL, the uncertainties for the macroscopic stellar parameters in the more restricted energy band are smaller due to reduced flux fluctuations. However, we decided to report the results from the broader energy band since they encompass a more representative dataset, in terms of the source's emission, and try to account for systematic modelling uncertainties that we did not explore in detail. 

For completeness, we did another Bayesian analysis also for the 3--8\,keV energy range with 16 bins/phase. We obtained that the main effect of decreasing the binning was the increase in the uncertainties of the observables with respect to the case with 50 bins/phase. Median values only changed slightly. For instance, radius uncertainties at $1\sigma$ CL were around 1.3\,km, while mass uncertainties were around 0.4\,$M_{\odot}$. Larger uncertainties are very much expected due to the reduction of data.

\subsection{Future improvements}

Our work represents an initial step towards systematic studies of the light-curve predictions of isolated NSs. To obtain more accurate NS parameter extractions, further improvements to the modelling and data reduction are needed. In future studies, we plan to 
(i) investigate atmospheric models that consider a wider range of elements (beyond hydrogen, oxygen and carbon) and magnetic field orientations (e.g., cases where the magnetic field is not parallel to the NS surface's normal), (ii) examine a more diverse range of hot spot geometries, beyond circular ones, (iii) do not ignore the orientation of magnetic field lines with the surface's normal, (iv) enlarge the magnetic field range in the NS atmosphere tables to also cover supercritical fields, (v) implement the beaming function self-consistently in our models, and (vi) allow more parameters to vary when fitting the observations.

\section{Summary} 
\label{sec:summary}
             
X-ray pulse profile analysis of 3XMM~J1852+0033 suggests that it is a low-B ($\sim 10^{11}-10^{12}$ G) magnetar with a large mass ($\sim 2.0-2.2\,M_{\odot}$) and a carbon atmosphere. Diffuse nuclear burning of hydrogen/helium would be a possible mechanism for the latter.
Posterior $M(R)$ distributions favor intermediate stiffness EOSs for purely hadronic stars. If exotic (e.g., quark) phases are present, phase transitions in which energy density jumps are small enough (weak phase transitions) are favored. Our best-fitting models feature three small hot spots, which suggest a multipolar magnetic field structure. The estimated magnetic field strength is subcritical (i.e., $B<4.4\times 10^{13}$ G). The above is in agreement with magnetars if their surfaces have been buried and their fields are just re-emerging. Finally, our results can be improved if better time resolution data is available, so future observations with 3XMM~J1852+0033 as the main target are greatly encouraged.

\appendix

\section{Pulse Profile model}
\label{ap:A}

Following our previous approach~(see \cite{2020ApJ...889..165D} for details), we consider an observer at $r\rightarrow \infty$ and a photon that arises from the stellar surface at $dS=R^2 \sin\theta d\theta d\phi$, where $R$ is the stellar radius, making an angle $\alpha$ with the local normal to the surface ($ 0 \leq \alpha \leq \pi/2$). The photon path is bent by an additional angle $\beta$ owing to the spacetime curvature, and the effective emission angle as seen by the observer is  $\psi=\alpha+\beta$. The geometry is symmetric relative to $\phi$. \cite{2002ApJ...566L..85B} has shown that a simple approximate formula can be used to relate the emission angle $\alpha$ to $\theta$:
\begin{equation}\label{Beloborodov_approximation}
    1 -\cos\alpha = (1-\cos\theta)\left(1-\frac{R_s}{R}\right),
\end{equation}
where $R_s=2GM/c^2$ is the Schwarzschild radius, $G$ is the gravitational constant and $M$ is the stellar mass. We note that Eq.~(\ref{Beloborodov_approximation}) is a very good approximation for $R>3R_s$ since it typically leads to very small errors ($\lesssim 1\%$). For the range of masses and corresponding radii of interest here, errors would be up to a few percent.

We assume that the stellar emission follows a local atmospheric spectrum, and that the observed flux comes mainly from  hot spots. The flux from the stellar photosphere is also considered, which is an improvement relative to  \cite{2020ApJ...889..165D} that only consider the emission from hot spots.

The intensity $I_\nu(T)$ at a given point on the stellar surface ($\nu$ is the photon frequency) depends on the temperature $T$ there, but it could also be related to an ``out of spot" atmospheric region. The flux is proportional to the visible area of the emitting region ($S_V$) plus a relativistic correction, and it is given by \citep{2002ApJ...566L..85B,2013ApJ...768..147T}

\begin{align}\label{FluxTurolla}
    F_\nu &=  \left(1-\frac{R_s}{R}\right)  \int_{S_V} I_\nu(T) \cos\alpha \frac{d\cos\alpha}{d(\cos\theta)}ds \ ,
 \end{align}

The model calculates the absolute count rates observed at  an infinite distance. 

In polar coordinates, the circular hot spot has its center at $\theta_0$ and a semi-aperture $\theta_c$. The spot is bound by the function $\phi_b(\theta)$, where $0\leq \phi_b \leq \pi$, given by \citep{2013ApJ...768..147T}

\begin{equation}
    \phi_b = \left[ \frac{\cos\theta_c - \cos\theta_0 \cos\theta }{\sin\theta_0 \sin\theta} \right] \ .
\end{equation}

Only the visible part of the star must be considered, therefore the spot must be also limited by a constant $\theta_F$ defined by
\begin{equation}\label{thetaF}
    \theta_{F}=\arccos{\left( 1- \frac{c^2 R}{2 G M} \right)^{-1}}\ .
\end{equation}
For a given bending angle $\beta$, that diverges from the emission angle $\alpha$, $\theta_F$ occurs for the maximum emission $\alpha$, i.e. $\alpha=\pi/2$ [see Fig.~1 of \cite{2020ApJ...889..165D}]. In Newtonian gravity, where $\beta=0$, the maximum visible angle is $\theta_F=\pi/2$, meaning that half of the stellar surface is visible. However, for a relativistic star, $\theta_F>\pi/2$.

By using the approximation given by Eq.~\ref{Beloborodov_approximation} into Eq.~\ref{FluxTurolla}, and integrating in $\phi$, one obtains
\begin{align}
      F_\nu &= 2  \left(1-\frac{R_s}{R}\right)^2  \int_{\theta_{min}}^{\theta_{max}} I_\nu(T,\theta) A(\theta)\sin\theta\ \phi_b(\theta) d\theta \ ,
\end{align}
where
$I_{\nu}=I_{\nu}(T)\cos\theta$, which takes into account non-isotropic emission \citep{2001ApJ...559..346D} (e.g., due to the magnetic field lines), and
\begin{align}
A(\theta) &= \left[ \frac{R_S}{R} + (1-\frac{R_S}{R})\cos\theta \right], \\
      \theta_{\text{min}} &= \text{min}[\text{max}(0,\theta_0-\theta_c),\theta_F]\ ,\\
      \theta_{\text{max}} &= \text{min}[\theta_0+\theta_c,\theta_F]\ .
\end{align}

The total flux produced by $N_\sigma$ spots, where the $\sigma$-th spot has a semi-aperture $\theta_{c\sigma}$ and a temperature $T_\sigma$, can be calculated by adding up each contribution, and so we have
\begin{eqnarray}\label{TotalFlux}
    F_\nu^{TOT} &=&  \sum_\sigma  F_\nu(T_\sigma)\ .
\end{eqnarray}
The pulse profile in a given energy band $[\nu_1,\nu_2]$ for a given spot $\sigma$ is

\begin{equation}
    F_\sigma(\nu_1,\nu_2)= \int_{\nu_1}^{\nu_2} F_\nu(T_\sigma) d\nu \ .
\end{equation}
Therefore, one can rewrite Eq.~(\ref{TotalFlux}) for a given energy band, and it becomes
\begin{eqnarray}\label{TotalFluxBand}
    F^{TOT}  = \sum_\sigma F_\sigma(\nu_1,\nu_2)  \ .
\end{eqnarray}

We define the Line of Sight (LOS) as the unity vector $\hat{\mathbf l}$ pointing from the neutron stellar center to the observer in the infinity. Also, we define by $\hat{\mathbf r}$ the unit vector parallel to the rotation axis of the star, whose angular velocity is $\Omega=2\pi/P$. 

We take $i$ as the angle between the LOS and the rotation axis ($\cos i=\hat{\mathbf r}\cdot\hat{\mathbf l}$). As the star rotates, the polar coordinate of the spot's center, $\theta_0$, changes. Let $\gamma(t)=\Omega t$ be the star's rotational phase. Thus, from a geometrical reasoning we have that
\begin{equation}\label{thetat}
    \cos\theta_0(t)=\cos i\cos \theta -\sin i\sin \theta\cos\gamma(t) \ ,
\end{equation}
where we have taken that $i$ and $\theta$ do not change with time.

\section{Atmosphere model table}
\label{ap:Atm}

Table \ref{tab:tabela_atm} gives the parameters for which the Nsmaxg model is tabulated. The energy range for the model is 0.05--10 keV (with 117 tabulated points), and it assumes hydrogen, carbon, or oxygen as the main atmospheric constituent. The magnetic field strength, local gravity, and temperature are the key inputs for computing the 
flux. Note that the atmosphere tables assume that the flux emerges 
at a zero angle with respect to the normal to the star surface. The number of tabulated points for $\log T_{eff}$ varies between 12 and 14 for a given $B$, surface gravitational field and energy value. For further details, see \citep{Ho_2008,2014IAUS..302..435H}.

\begin{table}
\centering
\caption{Parameters of the Nsmaxg model that are tabulated in XSPEC.}
\label{tab:tabela_atm}
\begin{tabular}{ccccccc}
\hline
Element & $B$ ($10^{12}$ G)  & $\log g$ (cm/s$^2$) & $\log T_{eff}$  \\ \hline
H & 0.01  & 2.4 & 5.5 - 6.7  \\
H & 0.04  & 2.4 & 5.5 - 6.7  \\
H & 0.07  & 2.4 & 5.5 - 6.7  \\
H & 0.1  & 2.4 & 5.5 - 6.7  \\
H & 1.0 & 0.4 - 2.5 & 5.5 - 6.8  \\
H & 2.0  & 2.4 & 5.5 - 6.8  \\
H & 4.0  & 2.4 & 5.5 - 6.8  \\
H & 7.0  & 2.4 & 5.5 - 6.8  \\
H & 10.0  & 2.4 & 5.6 - 6.8  \\
H & 30.0 & 2.4 & 5.7 - 6.8  \\
H & 1.26  & 1.6 & 5.5 - 6.8  \\
H & 7.0  & 1.6 & 5.5 - 6.8  \\
O &	1.0 &	2.4&	5.8 - 6.9	 \\
O &	10.0 &	2.4&	5.8 - 6.9	 \\
C & 1.0  & 2.4 & 5.8 - 6.9  \\
C & 10.0  & 2.4 & 5.8 - 6.9  \\
\hline
\end{tabular}
\end{table}
\vspace{1.0cm}
We thank the anonymous referee for the relevant suggestions which helped improve this work.
R.C.R.L. acknowledges the support of Fundação de Amparo à Pesquisa e Inovação do Estado de Santa Catarina (FAPESC) under grant No. 2021TR912. J.G.C. is grateful for the support of FAPES (1020/2022, 1081/2022, 976/2022, 332/2023), CNPq (311758/2021-5), FAPESP (2021/01089-1), and NAPI Fenômenos Extremos do Universo of Fundação Araucária. R.C.N. thanks the CNPq for partial financial support under the project No. 304306/2022-3 and the FAPERGS for partial financial support under the project No. 23/2551-0000848-3. P.E.S. acknowledges PCI/INPE/CNPq for financial support under grant No. 300320/2022-1. J.P.P. is grateful to FAPES under grant N. 04/2022. J.P.P., M.B., P.H. and J.L.Z. gratefully acknowledge the financial support of the National Science Center Poland grants no. 2016/22/E/ST9/00037 and 2018/29/B/ST9/02013. C.V.R. thanks the support from CNPq (310930/2021-9).
J.C.N.A. thanks CNPq (308367/2019-7) for partial financial support.

\bibliographystyle{elsarticle-harv} 
\bibliography{biblio}

\end{document}